\pdfoutput=1 %arXiv admin added this line
\documentclass[doublecol]{epl2} 
\usepackage{graphicx, epsfig, color}
\usepackage{subfigure}
\usepackage[english]{babel}
\usepackage[utf8]{inputenc}
\usepackage{cite}

\title{The backbone of the climate network}
\author{J. F. Donges\inst{1,2}\thanks{E-mail: \email{donges@pik-potsdam.de}} \and Y. Zou\inst{1} \and N. Marwan\inst{1} \and J. Kurths\inst{1,2}}
\shortauthor{J. F. Donges \etal}

\institute{                    
  \inst{1} Potsdam Institute for Climate Impact Research - P.O. Box 60 12 03, 14412 Potsdam, Germany\\
  \inst{2} Department of Physics, Humboldt University - Newtonstr. 15, 12489 Berlin, Germany
}

\abstract{
We propose a method to reconstruct and analyze a complex network from data generated by a spatio-temporal dynamical system, relying on the nonlinear mutual information of time series analysis and betweenness centrality of complex network theory. We show, that this approach reveals a rich internal structure in complex climate networks constructed from reanalysis and model surface air temperature data. Our novel method uncovers peculiar wave-like structures of high energy flow, that we relate to global surface ocean currents. This points to a major role of the oceanic surface circulation in coupling and stabilizing the global temperature field in the long term mean (140 years for the model run and 60 years for reanalysis data). We find that these results cannot be obtained using classical linear methods of multivariate data analysis, and have ensured their robustness by intensive significance testing.}

\begin{document}

\maketitle

%%%
%%%  Section: Introduction
%%%
\section{Introduction}

In the last decade, the complex network paradigm has proven to be a fruitful tool for the investigation of complex systems in various areas of science, \emph{e.g.}, the internet and world wide web in computer science, food webs, gene expression and neural networks in biology, and citation networks in social science \cite{BasicNetworkReviews}. The intricate interplay between the structure and dynamics of real networks has received considerable attention \cite{Boccaletti2006}. Particularly, synchronization arising by the transfer of dynamical information in complex network topologies has been studied intensively \cite{Arenas200893}. The application of complex network theory to climate science is a young field, where only few studies have been reported recently \cite{TsonisBasicClimNet, tsonis2008tap, yamasakigozol2008, Donner2008nts,donges2009epjst}. The vertices of a climate network are identified with the spatial grid points of an underlying global climate data set. Edges are added between pairs of vertices depending on the degree of statistical interdependence between the corresponding pairs of anomaly time series taken from the climate data set. Climate networks enable novel insights into the topology and dynamics of the climate system over many spatial scales ranging from local properties as the number of first neighbors of a vertex $v$ (the degree centrality $k_v$) to global network measures such as the clustering coefficient or the average path length. The local degree centrality and related measures have been used to identify supernodes (regions of high degree centrality) and to associate them to known dynamical interrelations in the atmosphere, called teleconnection patterns, most notably the North Atlantic Oscillation (NAO) \cite{TsonisBasicClimNet}. On the global scale, climate networks were found to possess ``small-world'' properties due to long range connections (edges linking geographically very distant vertices), that stabilize the climate system and enhance the energy and information transfer within it \cite{TsonisBasicClimNet}. By studying the prevalence of long range connections in El Ni\~no and La Ni\~na climate networks \cite{tsonis2008tap} and the time dependence of the number of stable edges \cite{yamasakigozol2008}, it has been shown very recently, that the El Ni\~no-Southern Oscillation (ENSO) has a strong impact on the stability of the climate system.

Until now, researchers have used the linear cross-correlation function of pairs of anomaly time series to quantify the degree of statistical interdependence between different spatial regions. But the highly nonlinear processes at work in the climate system call for the application of nonlinear methods to obtain more reliable results. Here we also use mutual information \cite{kantz2004nts} to construct climate networks allowing to capture linear and nonlinear relationships between time series \cite{donges2009epjst}. Furthermore we use a measure of vertex centrality, betweenness (BC), that is defined locally but takes into account global topological information. Combining these two techniques, we uncover peculiar wave-like structures in the BC fields of climate networks constructed from monthly averaged reanalysis and atmosphere-ocean coupled general circulation model (AOGCM) surface air temperature (SAT) data. Akin to the homonymous data highways of the internet, these BC structures form the backbone of the SAT network, bundling most of the energy flow between remote regions. Some major features of the backbone appear to be closely related to surface ocean currents pointing to an essential role of the oceanic surface circulation in stabilizing the climate system by promoting the global flow of energy, mainly in the form of heat. Note that these insights are conceptually new and cannot be obtained using classical methods of climatology such as principal component analysis (PCA) or singular spectrum analysis (SSA) of anomaly fields \cite{ClimateStatistics}, because these are by design local in a network sense and are not suitable to study local flow measures depending on a global network topology. We have performed intensive statistical tests with various types of surrogates to ensure the robustness of our results. The methodology developed in this letter has the potential to be universally applicable to extract the energy, matter or information flow structure in any spatially extended dynamical system from observations taken from the real world, experiments and simulations. Our results are hence of interest for several branches of physics as well as various applications, \emph{e.g.}, fluid dynamics (turbulence), plasma physics, biological physics (population dynamics, neural networks, cell models). As demonstrated by its application to the climate system, our method is particularly relevant for the analysis of systems with highly heterogeneous boundary conditions and forcing fields, that are found frequently in nature.

%%%
%%% Section: Data
%%%
\section{Data}

We utilize the monthly averaged global SAT field to construct climate networks, that allows to directly capture the complex dynamics on the interface between ocean and atmosphere due to heat exchange and other local processes. SAT therefore enables us to study atmospheric as well as oceanic dynamics using the same climate network. We use reanalysis data provided by the National Center for Environmental Prediction/National Center for Atmospheric Research (NCEP/NCAR) \cite{kistler2001nny} and model output from the World Climate Research Programme's (WCRP's) Coupled Model Intercomparison Project phase 3 (CMIP3) multi-model data set \cite{meehl2007wcm}. For optimal comparability with the reanalysis data, we choose a 20th century reference run (20c3m, as defined in the IPCC AR4) by the Hadley Centre HadCM3 model. A data set consists of a regular spatio-temporal grid with time series $x_i(t)$ associated to every spatial grid point $i$ at latitude $\lambda_i$ and longitude $\phi_i$. Start and end dates, length of time series $\mathcal{T}$, latitudinal resolution $\Delta\lambda$, longitudinal resolution $\Delta\phi$ and the number of vertices of the corresponding global climate network $N$ are given in table~\ref{GlobalDatasetsTable}.

%%%
%%% Table containing properties of the data sets used
%%%
\begin{table}[t]%[H] add [H] placement to break table across pages
\caption{\label{GlobalDatasetsTable}Properties of global surface air temperature data sets}
\begin{center}
\begin{tabular}{l|c|c}
 & NCEP/NCAR & HadCM3\\
 \hline
Period & 01/1948 - 12/2007 & 01/1860 - 12/1999\\
$\mathcal{T}$ [months] & 720 & 1,680\\
$\Delta\lambda$ [$^\circ$] & 2.5 & 2.5\\
$\Delta\phi$ [$^\circ$] & 2.5 & 3.75\\
$N$ & 10,224 & 6,816\\
\end{tabular}
\end{center}
\end{table}

%%%
%%% Section: Methodology
%%%
\section{Methodology}

%%%
%%% Phase averaging
%%%
(i) To minimize the bias introduced by the external solar forcing common to all time series in the data set, we calculate anomaly time series $a_i(t)$ from the $x_i(t)$, \emph{i.e.}, remove the mean annual cycle by phase averaging. Up to this point, we follow the method used previously by \cite{yamasakigozol2008, tsonis2008tap}. It is known, that the annual cycle induces higher order effects such as seasonal variability of anomaly time series variance. We find that using only data from a particular season to avoid biases due to this effect does not alter our results substantially, so that we choose to use the whole data set for a more accurate evaluation of interdependence. Furthermore we normalize the anomaly time series to zero mean and unit variance.

%%%
%%% Mutual information
%%%
(ii) Mutual information (MI) is a measure from information theory, that can be interpreted as the excess amount of information generated by falsely assuming the two time series $a_i$ and $a_j$ to be independent, and is able to detect linear as well as nonlinear relationships \cite{kantz2004nts}. MI can be estimated using 

\begin{equation}
M_{ij} = \sum_{\mu \nu} p_{ij}(\mu, \nu) \log \frac{p_{ij}(\mu, \nu)}{p_i(\mu) p_j(\nu)}, \label{MIFormula}
\end{equation}

\noindent where $p_i(\mu)$ is the probability density function (PDF) of the time series $a_i$, and $p_{ij}(\mu, \nu)$ is the joint PDF of a pair $(a_i,a_j)$. By definition, $M_{ij}$ is symmetric, so that $M_{ij} = M_{ji}$. Note that in principle, one can evaluate a time delayed MI \cite{kantz2004nts}. This is appropriate when studying climate networks on smaller time scales using data sets with (sub-)diurnal resolution \cite{yamasakigozol2008}. However, in the present work, we intend to study long term structural properties of the climate system on a scale of $\mathcal{O}(10^2)$ years using monthly averaged data. Most physical mechanisms of global information transfer in the SAT field such as travelling Rossby waves, heat exchange between ocean and atmosphere or the advection of heat by surface currents in the ocean act at time scales of less than one month. Therefore, it is reasonable to calculate {MI} at zero lag between anomaly time series.

%%%
%%% Adjacency matrix
%%%
(iii) We now construct the climate network by thresholding the {MI} matrix $M_{ij}$, \emph{i.e.}, only pairs of vertices $(i,j)$ that satisfy $M_{ij}>\tau$ are regarded as linked, where $\tau$ is the threshold. Using the Heaviside function $\Theta(x)$, the adjacency matrix $A_{ij}$ of the climate network is given by {$A_{ij} = \Theta \left( M_{ij} - \tau \right) - \delta_{ij}$, where $\delta_{ij}$ is the Kronecker delta.} Note that $A_{ij}$ inherits its symmetry from $M_{ij}$ and the resulting climate network is an undirected and unweighted simple graph. One could construct a network with edges $(i,j)$ weighted by $M_{ij}$. At this stage, we keep our method simple by studying an unweighted network. We suggest that weight information could help to identify the backbone structure even more clearly, however, would not alter our conclusions below, because we use only a small number of edges with high {MI} that dominate the network.

%%%
%%% Edge density
%%%
(iv) {We find, that network characteristics, such as {BC}, clustering coefficient and average path length, are dependent on the choice of the threshold $\tau$.} When comparing climate networks constructed from AOGCM and reanalysis data, it is consequently more meaningful to constrain the edge density $\rho = 2E /  (N(N-1))$, where $E$ gives the total number of edges, than to fix $\tau$ as it was done in all earlier works \cite{TsonisBasicClimNet, tsonis2008tap, yamasakigozol2008}. The threshold $\tau = \tau(\rho)$ is thus chosen to yield a prescribed edge density $\rho$. The PDF of {MI} $P(M)$ over all pairs $M_{ij}$ is found to have a connected support, so that the edge density function $\rho(\tau) = \int_{\tau}^{\infty}\upd M P(M)$ is strictly monotonic decreasing with $\tau$ and induces a one to one correspondence between $\tau$ and $\rho$. Note that the backbone of the climate network is most clearly observed at small $\rho$ with corresponding large threshold $\tau$, that is very unlikely to be exceeded by chance, as we reassured using significance tests based on randomly shuffled time series, Fourier surrogates \cite{kantz2004nts} and twin surrogates \cite{thiel2006tst}. {We fix the edge density at $\rho = 0.005$, resulting in thresholds of $\tau_1 = 0.398$ for the HadCM3 data and $\tau_2 = 0.624$ for the reanalysis data. The remaining $0.5\%$ of all possible edges correspond to statistically significant and robust relationships (see below). In concordance with this observation, we find that small variations of $\rho$ from the chosen value do not alter the backbone structure significantly. The remaining edges are distributed heterogenously as they attach preferentially to pronounced supernodes, their range extending from local to global (teleconnections), which is consistent with earlier works \cite{TsonisBasicClimNet,donges2009epjst}.}

%%%
%%% Betweenness
%%%
(v) Having constructed a climate network, we can finally quantify the importance of a small part of the earth's surface (represented by a single vertex $v$) for the {global flow of energy} within the SAT field, that gives rise to the pairwise dynamical interdependencies measured by {MI}. {Vertices play distinct roles in the energy transmission throughout the network, some of them show a higher capability as compared to others.} This capability can be quantified by the betweenness $BC_v$. Assume that energy travels through the network on shortest paths. {We then regard a vertex $v$ to be important for the energy transport in the network, if it is traversed by a large number of all existing shortest paths \cite{freeman1979csn}. The betweenness centrality is defined as

\begin{equation}
BC_v = \sum_{i,j\neq v}^N \frac{\sigma_{ij}(v)}{\sigma_{ij}}, \label{eq:betweenness}
\end{equation}

\noindent $\forall v \in (1,\dots,N)$, where $\sigma_{ij}(v)$ gives the number of shortest paths from $i$ to $j$, that include $v$. Here the contribution of shortest paths is weighted by their respective multiplicity $\sigma_{ij}$.} {Because the shortest paths considered contain only edges corresponding to pairs of highly dynamically interrelated time series, {BC} can be interpreted as a local measure of dynamical information flow. Since we use it to analyze a temperature field we nevertheless prefer to view {BC} more fundamentally as a measure of the flow of energy, mainly in the form of heat.} {{BC} is conceptually distinct from other commonly used vertex centrality measures, \emph{e.g.}, degree and closeness centrality, and hence enables us to uncover interesting novel structural features of climate networks \cite{donges2009epjst}.}

%%%
%%% Results
%%%
\section{Results}

%%%
%%% Main figure diplaying the central BC result
%%%
\begin{figure}[t]
\centering
\subfigure{
\includegraphics[width=0.97\columnwidth, viewport=35 55 716 537, clip]{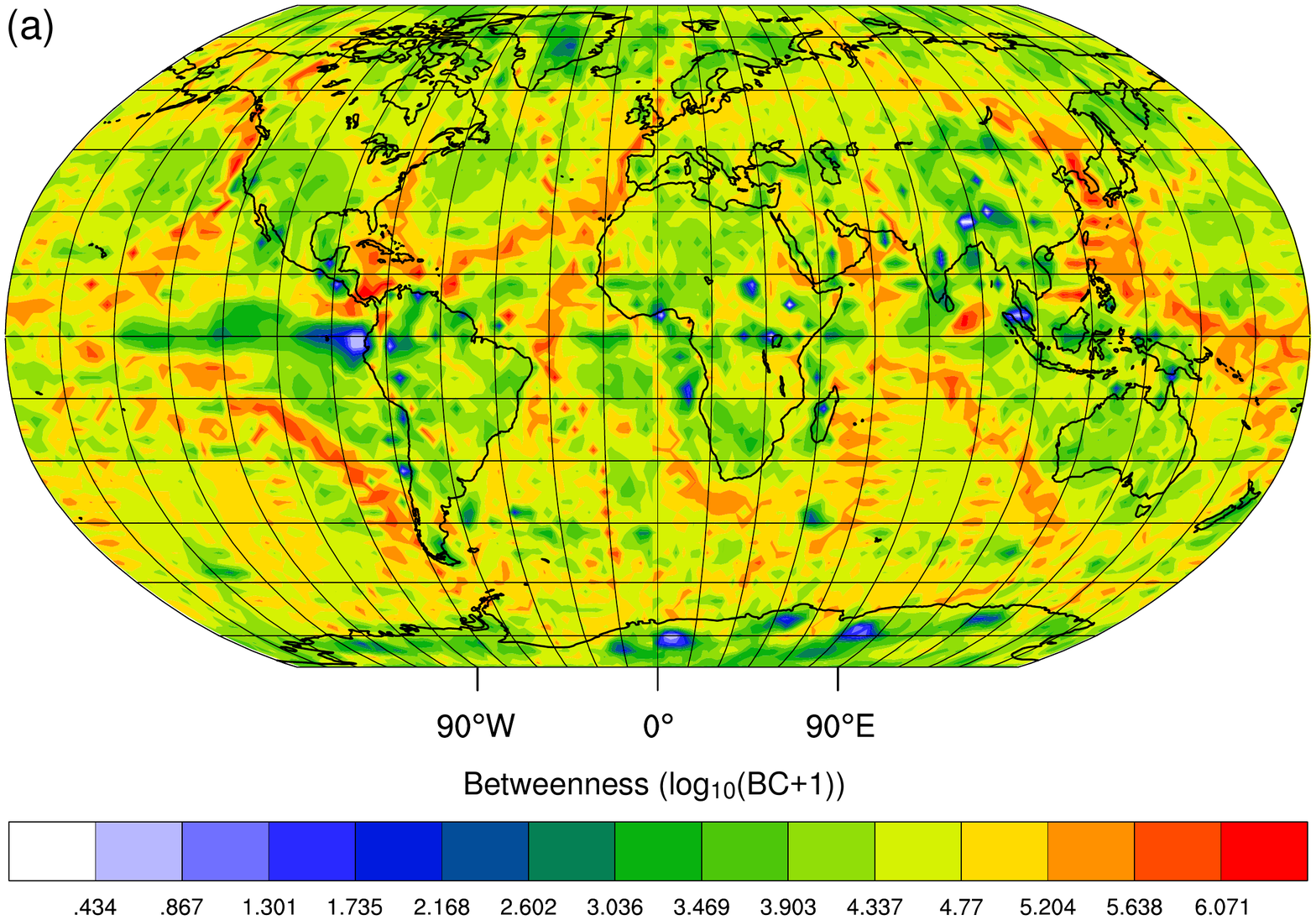}
\label{fig:OceanCurrentsSub3}
}
\subfigure{
\includegraphics[width=0.97\columnwidth, viewport=35 55 716 537, clip]{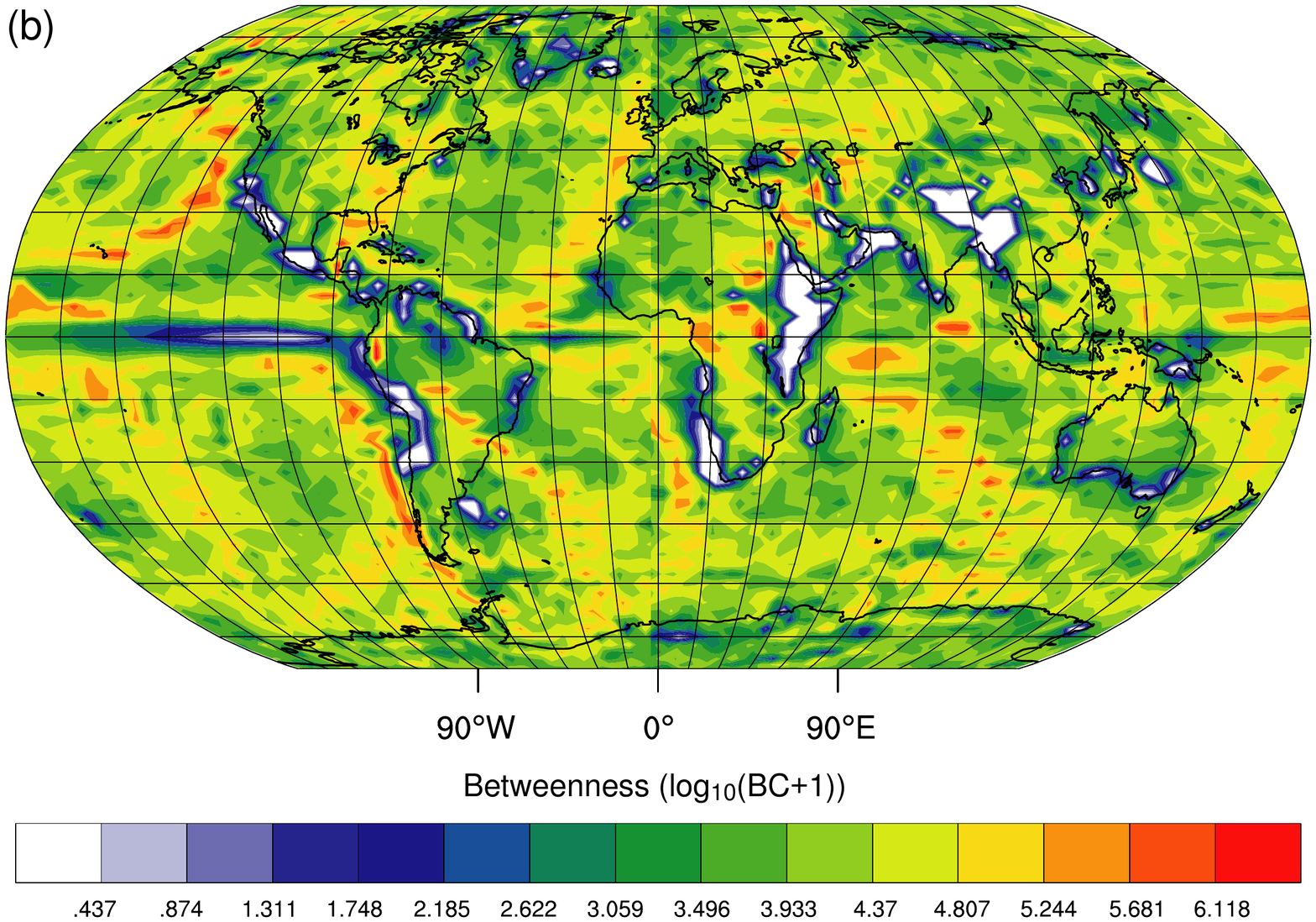}
\label{fig:OceanCurrentsSub2}
}
\subfigure{
\includegraphics[width=0.97\columnwidth,viewport=37 48 666 400, clip]{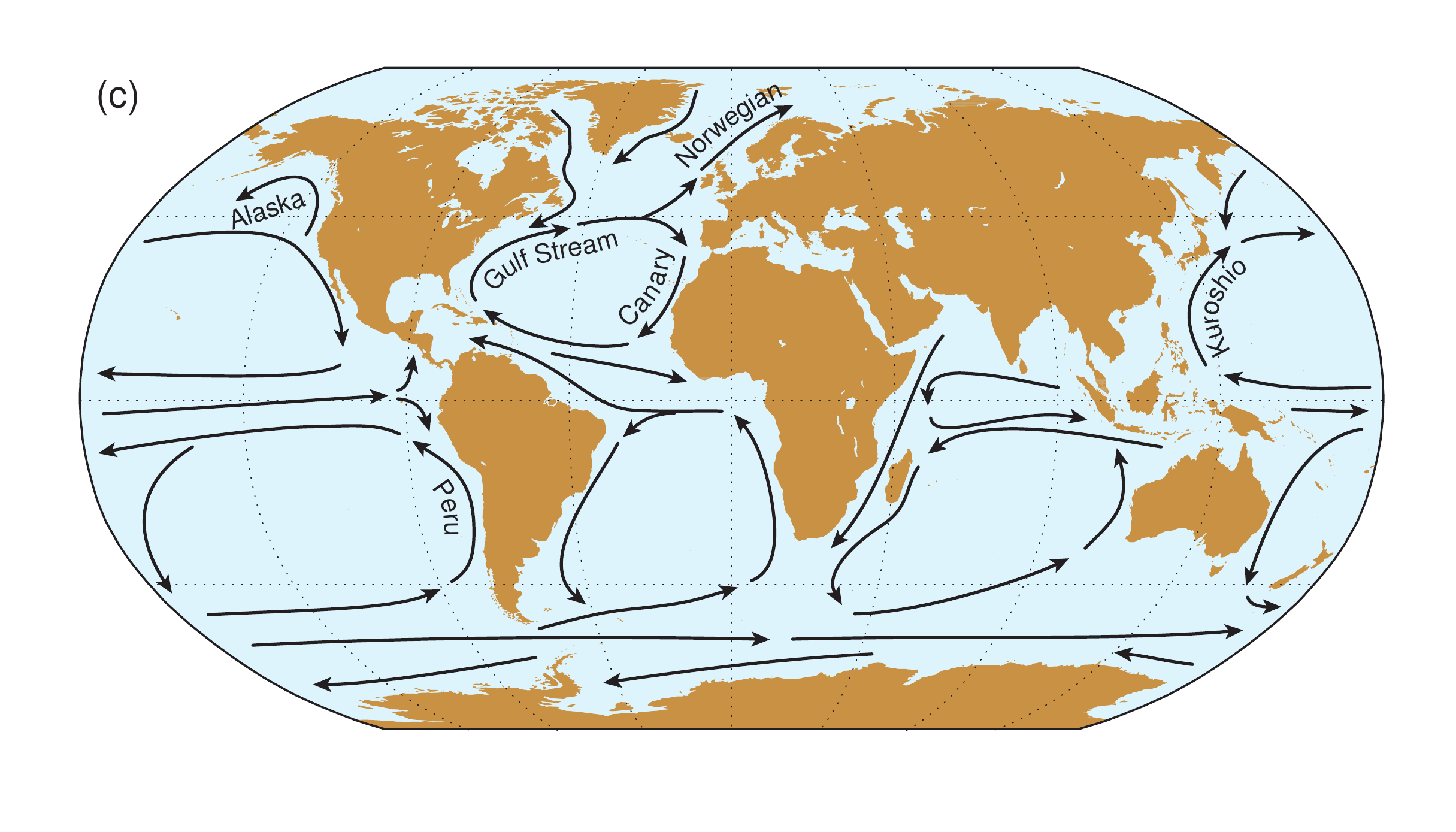}
\label{fig:OceanCurrentsSub1}
}
\caption{\label{fig:OceanCurrents}a) {BC} for the NCEP/NCAR reanalysis SAT {MI} network, and b) for the HadCM3 SAT network. Both networks are constructed at edge density $\rho=0.005$ using {MI}. c) A schematic map of global surface ocean currents, after \cite{Pidwirny2006}. Note that some features of the backbone in a) and b) correspond closely to ocean surface currents shown in c), \emph{e.g.}, the Alaska, Peru and Canary currents.}
\end{figure}

Following the method outlined above, we uncover peculiar wave-like structures of high {BC} in fields of both reanalysis and model SAT climate networks (fig.~\ref{fig:OceanCurrents}). In analogy with the internet, we call the network of these channels of high {energy} flow the backbone of the climate network. We observe that prominent mainly meridional features of the backbone tend to approach the equator tangentially, as one would expect from modes of the atmospheric and oceanic general circulation due to the vanishing coriolis force at the equator \cite{vallis2006aao}. There is also a qualitative agreement on the location of major backbone structures for both reanalysis (fig.~\ref{fig:OceanCurrentsSub3}) and model networks (fig.~\ref{fig:OceanCurrentsSub2}), \emph{e.g.}, the high {BC} channel over the Atlantic Ocean and the backbone structures over the eastern Pacific Ocean, both connecting the Arctic with the Antarctic. 

\begin{figure}[t]
\centering
\includegraphics[width=0.97\columnwidth, viewport=35 55 716 537, clip]{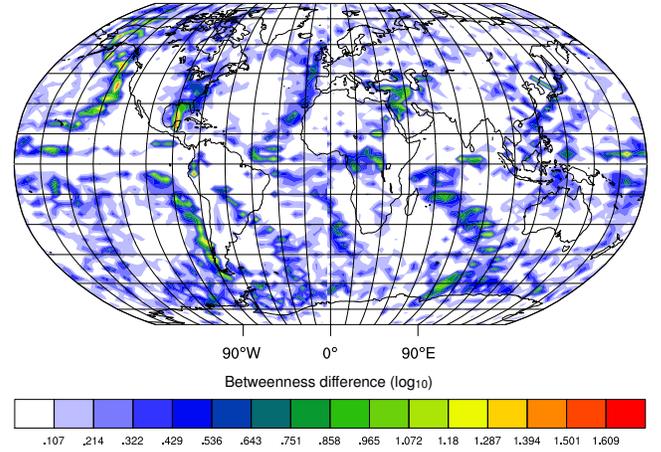}
\caption{\label{fig:BetweennessDifferenceField}Normalized difference field $\Delta BC_v = |BC^P_v - BC^M_v| / \sqrt{\left<BC^P_w\right>_w \left<BC^M_w\right>_w}$ of {BC} fields $BC^P_v$ and $BC^M_v$, respectively calculated from {PC} and {MI} HadCM3 SAT climate networks at $\rho=0.005$.}
\end{figure}

{The {BC} field of the MI climate networks considered here shows qualitative and quantitative deviations when compared to climate networks constructed using the linear Pearson correlation (PC) (fig.~\ref{fig:BetweennessDifferenceField}), while the backbone is clearly seen in both types of networks. The observed deviations could be partly due to a heterogeneous distribution of Shannon entropy among the $a_i(t)$, introducing a bias in the calculation of $M_{ij}$. Also it is well known that the nonlinear features of temperature dynamics might vary among reanalysis data sets, among climate models as well as between reanalysis and model data sets. Nevertheless we argue in \cite{donges2009epjst}, that some of the deviations in the {BC} field may be attributable to nonlinear physical processes in the climate system. Particularly, we find that edges corresponding to nonlinear statistical interrelationships are present in the MI climate network, which are excluded in the PC network at the same restrictive edge density level.}

%  Betweenness difference field figure

Note that the strongest backbone structures lie mainly over the ocean and avoid to cross the land in both model and reanalysis climate networks. Therefore a physical mechanism involving an atmosphere-ocean coupling might {contribute to the energy transport} in the SAT field measured by {BC}. Indeed, some of the strongest features found in the {NCEP/NCAR and HadCM3 {BC} fields} (fig.~\ref{fig:OceanCurrentsSub3} and \ref{fig:OceanCurrentsSub2}) resemble closely major surface ocean currents (fig.~\ref{fig:OceanCurrentsSub1}). For example, note the strong {BC} structures off the west coast of North and South America that resemble the Alaska and Peru current, and the backbone feature along the west coasts of Africa and Europe following the path of the Canary and Norwegian currents. These observations can be understood considering the strong coupling between sea surface temperature (SST) and SAT over the ocean via heat flux. Temperature anomalies in SST are advected by the surface ocean currents and transfered to the SAT field via heat flux coupling. Therefore, ocean currents provide a physical mechanism for the transport of {energy together with dynamical information} on localized linear structures over large distances. However, no clear traces of the strong western boundary currents (WBCs) such as the Gulf Stream or the Kuroshio are visible in the backbone structure (fig.~\ref{fig:OceanCurrentsSub2}). This might be due to the fact, that WBCs are much narrower than the eastern boundary currents discussed above \cite{vallis2006aao}, so that the effect of WBCs is not resolved by the grid underlying the HadCM3 climate network (see table~\ref{GlobalDatasetsTable}). {Note that using higher resolution reanalysis data (fig.~\ref{fig:OceanCurrentsSub3}) and SAT data taken from the AOGCM ECHAM5 \cite{meehl2007wcm}}, we find that our method does indeed detect WBCs. {Here it should be pointed out again that we are analyzing the SAT field, hence purely atmospheric effects, \emph{e.g.}, planetary waves, also contribute to the {BC} field and might explain some of its wave-like features, particularly over land.}

Backbone structures are not seen in fields of the complementary random walk betweenness \cite{RandomWalkBetweenness} which measures diffusive {flow} in a network. This further supports our argument that shortest path betweenness (BC) measures convective {energy flow} in a spatially extended network and is consistent with extremalization principles of physics, \emph{e.g.}, the Hamiltonian principle, interpreted within a graph theoretical framework.

%  SST-SAT Gradient figure
\begin{figure}[t]
\centering
\includegraphics[width=0.97\columnwidth]{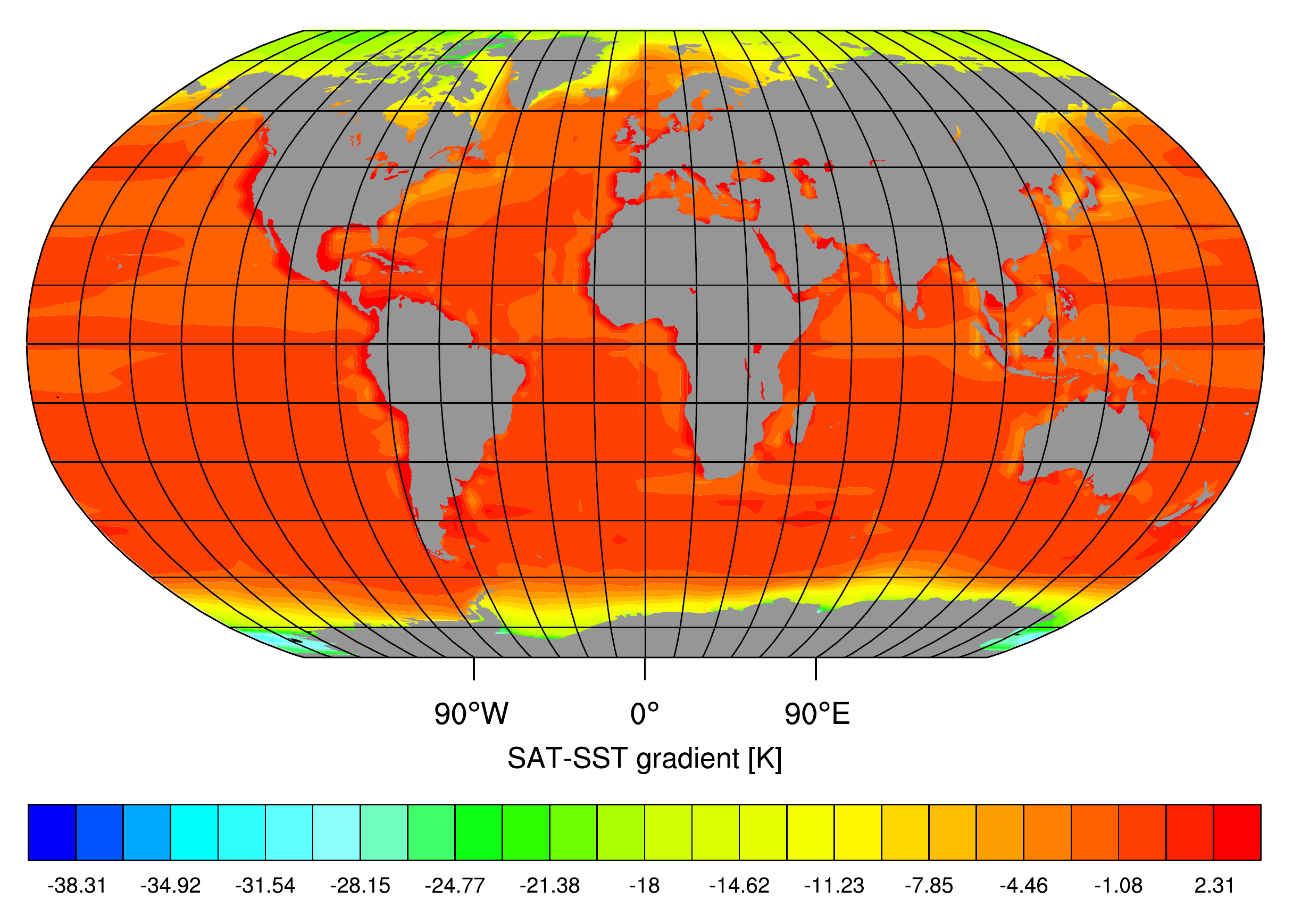}
\caption{\label{fig:SST-SAT-gradient}{The mean SAT-SST gradient field $\left<\Delta T_v(t)\right>_t = \left<SAT_v(t)\right>_t - \left<SST_v(t)\right>_t$ calculated from the HadCM3 SAT and SST data sets \cite{meehl2007wcm}, both taken from the 20c3m run.}}
\end{figure}

\begin{figure}[t]
\centering

\includegraphics[width=0.49\columnwidth, viewport=0 0 390 400, clip]{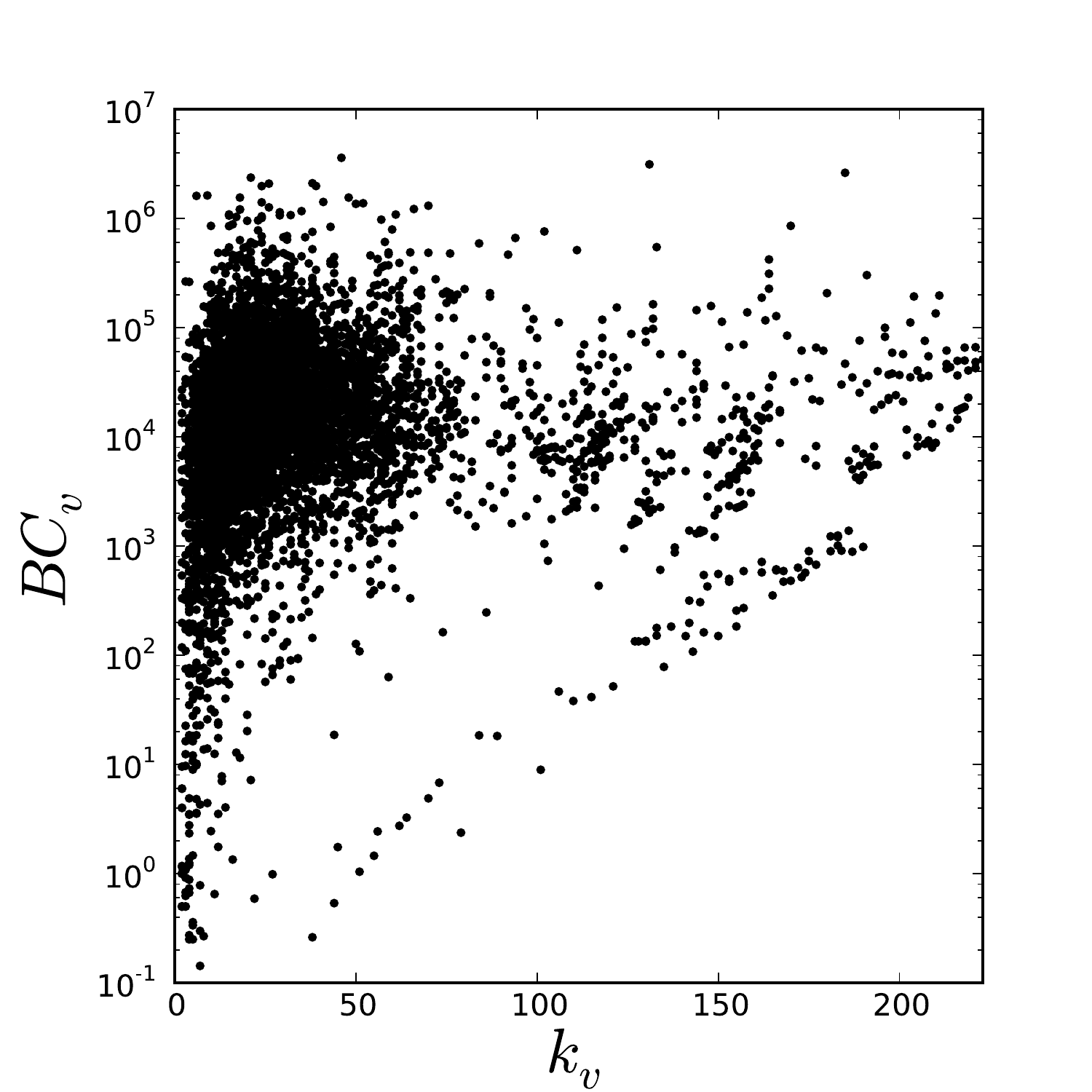}
\includegraphics[width=0.49\columnwidth, viewport=0 0 390 400, clip]{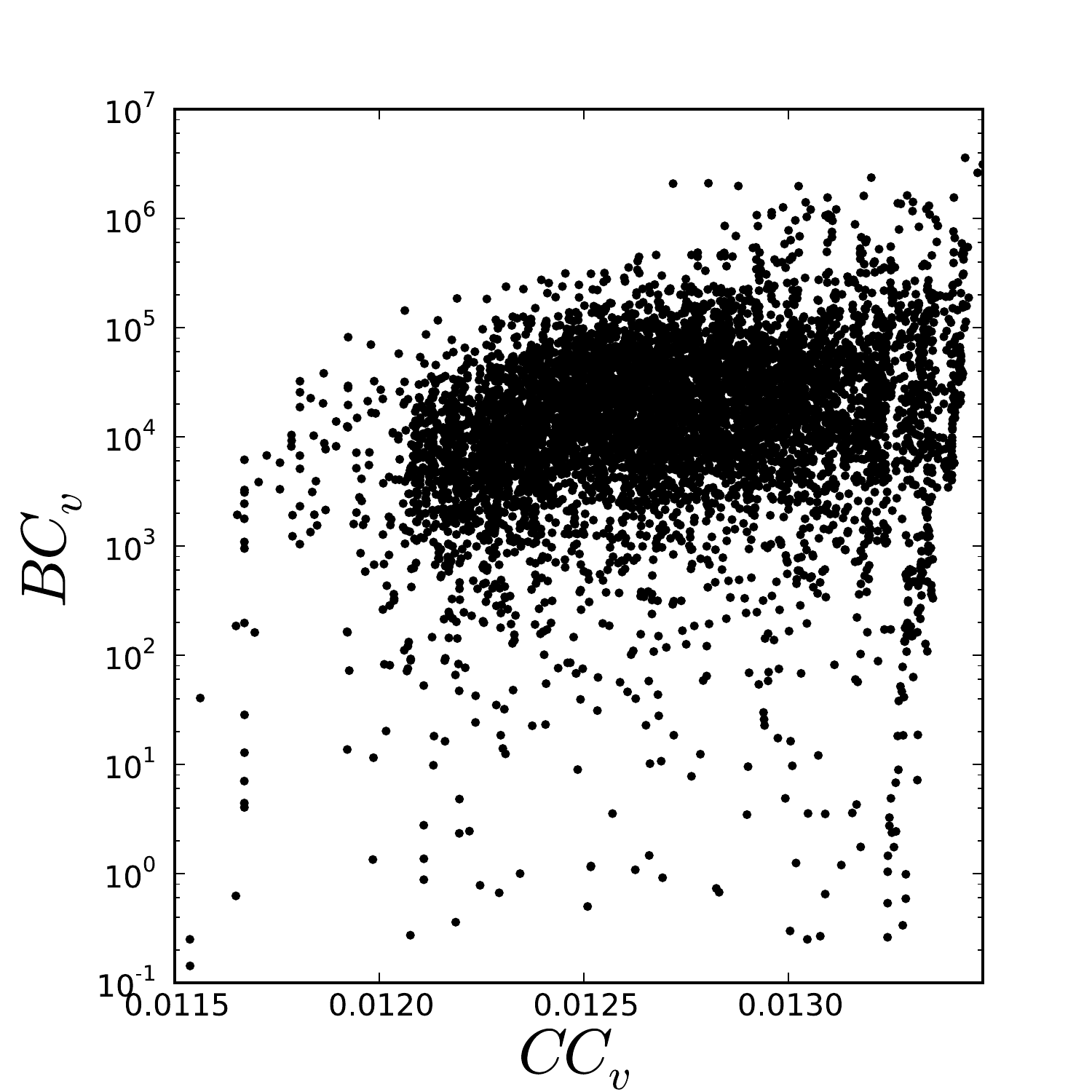}

\caption{\label{fig:ScatterPlots}{Scatter plots of betweenness $BC_v$ against (a) degree $k_v$ and (b) closeness ($CC_v$, \cite{freeman1979csn}) for the HadCM3 SAT {MI} climate network at $\rho=0.005$. Specifically, the Spearman's rank correlation coefficients of the centrality fields are $r_s(k_v,BC_v) = 0.25$ and $r_s(CC_v,BC_v) = 0.30$.}}
\end{figure}

To exclude the possibility that the observed backbone structures over the ocean might be simply due to local anomalies in the SST-SAT gradient caused by surface currents, we have calculated the gradient field from the model run that we used to construct the HadCM3 climate network, and found that the SST-SAT gradient and {BC} are not correlated (fig. \ref{fig:SST-SAT-gradient}). Furthermore, the backbone is neither seen in fields of degree nor closeness centrality {\cite{donges2009epjst,freeman1979csn}, while {BC} statistically shows some correlation with these centrality measures (fig. \ref{fig:ScatterPlots}). Nevertheless there is a notable tendency of high {BC} vertices to have a small degree, suggesting that they act as bottlenecks of energy flow in the network.} Therefore we conclude that the backbone structures observed in model and reanalysis networks are neither a trivial response to local anomalies in the SST-SAT gradient nor artifacts of chains of supernodes with high degree and closeness centrality.

{It is noteworthy that due to the given grids (table~\ref{GlobalDatasetsTable}), the vertex density on the earth's surface is maximum at the poles and decreases towards the equator. However, preliminary results using a density invariant generalization of {BC} suggest that our results are not appreciably impacted by this inhomogeneity.}

%%%
%%% Scatter plot figure
%%%

%%%
%%% Significance testing
%%%
\section{Significance testing}

%%%
%%% Significance testing figure
%%%
\begin{figure}[t]
\centering
\subfigure{
\includegraphics[width=0.97\columnwidth, viewport=35 55 716 537, clip]{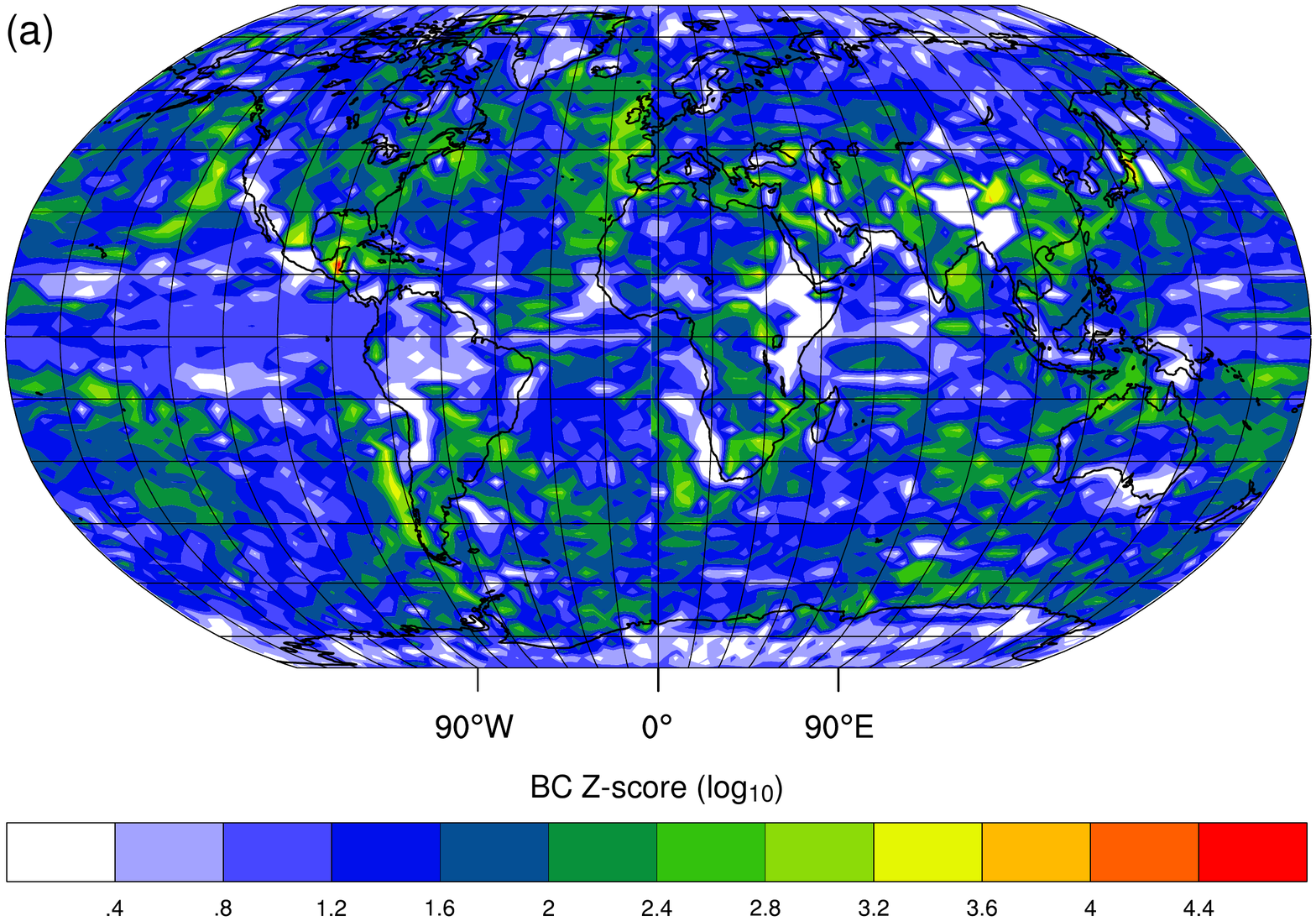}
\label{fig:SignificanceTesting1}
}
\subfigure{
\includegraphics[width=0.97\columnwidth, viewport=35 55 716 537, clip]{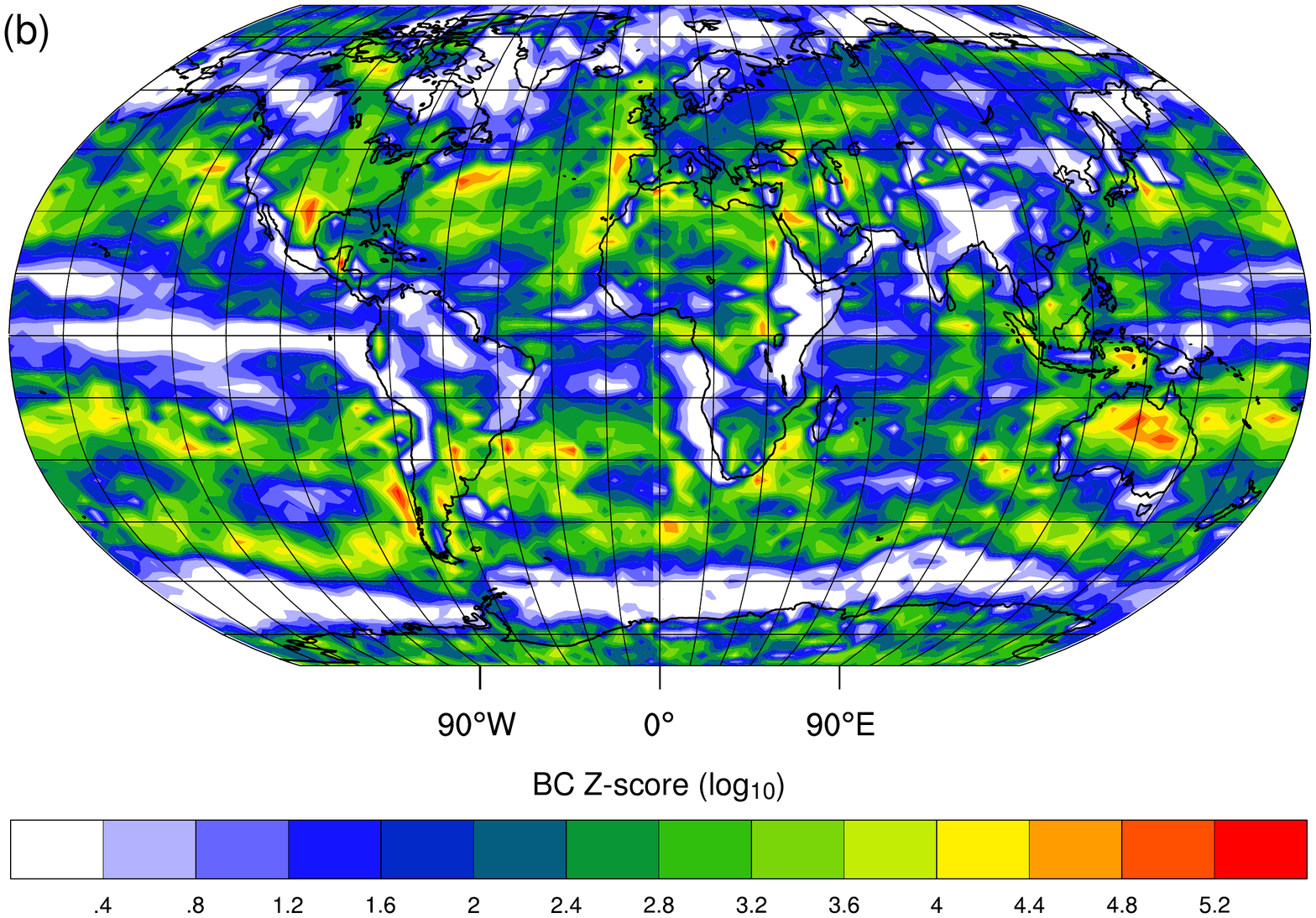}
\label{fig:SignificanceTesting2}
}
\caption{\label{fig:SignificanceTesting}Z-score field $Z_v= (BC_v - m(BC^r_v)) / \sigma(BC^r_v)$ of the {BC} field with respect to (a) a configuration model ensemble and (b) an ensemble based on twin surrogate data sets with both $n=100$ members calculated for the HadCM3 SAT {MI} climate network at $\rho=0.005$ \cite{zacharias2002plm}. $m(BC^r_v)$ and $\sigma(BC^r_v)$ denote the ensemble mean and standard deviation of {BC} at each vertex $v$, respectively. {If $|Z_v| \gg 1$, we can consider $BC_v$ to be significant with respect to the chosen network model. The backbone is recognizable with a large Z-score, indicating its statistical significance.}}
\end{figure}

To ensure the statistical robustness of our results on the network level, we test the null hypothesis, that the climate network is a random graph with a given degree sequence. Using the configuration model and a link switching method \cite{BasicNetworkReviews}, we generate a Monte Carlo ensemble of 100 networks, that have approximately the same degree sequence as the reconstructed climate network. We find that in sharp contrast to the reconstructed network, the ensemble mean {BC} sequence is highly correlated to the degree sequence, and does not display the backbone structures observed in the reconstructed climate network (fig.~\ref{fig:SignificanceTesting1}). Based on this evidence we reject the null hypothesis that the climate network is random and conclude, that the backbone is unlikely to be a trivial consequence of the degree sequence.

An alternative is to test the statistical robustness on the time series level and to develop the null hypothesis that the time series of the SAT data set are pairwise independent. Specifically, we generate 100 twin surrogates from the original time series at each grid point. {Compared to shuffled time series or Fourier surrogates, twin surrogates yield a higher test power since they approximately conserve all linear and nonlinear properties of the original single time series.} We then construct an ensemble of 100 networks from the resulting surrogate data sets, {again fixing the edge density at $\rho=0.005$, and compute the ensemble mean degree and betweenness sequences.} While interestingly, the ensemble mean degree sequence closely resembles the degree sequence of the climate network, the ensemble mean {BC} sequence is again highly correlated to the ensemble mean degree sequence and contains no backbone structures (fig.~\ref{fig:SignificanceTesting2}). Based on these observations, we reject the null hypothesis that the time series of the SAT data set are pairwise independent and infer, that the backbone indeed characterizes the intrinsic complex topology of dynamical interrelationships. We have performed these statistical tests for both the reanalysis and model climate networks and came to the same conclusions.

%%%
%%% Conclusions / Outlook
%%%
\section{Conclusions and outlook}

In summary, using mutual information from nonlinear time series analysis and betweenness from complex network theory, we have uncovered novel pathways of global {energy and dynamical information flow} in the climate system, that we call the backbone of the climate network. Two conceptually independent types of tests reveal that the backbone does not arise by chance and is not a trivial consequence of the degree centrality sequence studied in previous works on climate networks, but on the contrary represents a statistically significant feature of the underlying SAT data set. Surface ocean currents appear to play a major role in the {energy and information transfer} and hence in the dynamical stabilization of the climate system in the long term mean (140 years for the HadCM3 model run and 60 years for the reanalysis data). We observe similar backbone structures in AOGCM model output and reanalysis data giving confidence that the backbone is not a model artifact. It is important to realize that our complex network approach is an essential ingredient in the discovery of the backbone. The main advantage of betweenness is that it takes into account the global network topology of pairwise interrelationships between regions. However, the classical linear methods (PCA, SSA, etc. \cite{ClimateStatistics}) widely applied to disclose teleconnection patterns in climatology use information about next neighbors at each grid point, and are therefore only local within the complex network framework. Our method is promising to next study the impact of extreme events such as strong El Ni\~nos, extreme Monsoons or volcanic eruptions on the topology of climate networks. In the future it will thereby allow us to obtain new insights into the individual local signature of changes in the {energy and information flow structure} and stability of the climate system. Our method may also be valuable to illuminate differences in the backbone structure of different climate states in earth's history, \emph{e.g.}, during glacial and interglacial episodes, and to assess the impact of global warming on the stability of the climate system from a different perspective. More generally, we emphasize that the methodology introduced in this letter is universal in the sense, that it can be applied to study the structure of {energy, matter and information flow} within any spatially extended dynamical system.

%%%
%%% Acknowledgements
%%%
\acknowledgments

{We thank A. Bergner, G. Zamora-L\'{o}pez, A. Levermann and G. Schmidt for helpful discussions, and J. Heitzig for his valuable contributions.} JK acknowledges the SFB 555 (DFG) for financial support. JFD acknowledges the German National Academic Foundation.

%%%
%%%  Bibliography
%%%

\end{document}